\documentclass[prb, twocolumn, superscriptaddress, showpacs]{revtex4-1}

\usepackage{amsmath, amsthm, amssymb}
\usepackage{graphicx}
\usepackage{color}
\usepackage{bm}
\usepackage{hyperref, verbatim}

\newcommand{\bal}{\begin{align} }
\newcommand{\eal}{\end{align} }
\newcommand{\be}{\begin{equation} }
\newcommand{\ee}{\end{equation} }
\newcommand{\ba}{\begin{eqnarray} }
\newcommand{\ea}{\end{eqnarray} }
\newcommand{\ket}[1]{|#1\rangle}
\newcommand{\bra}[1]{\langle #1 |}

\begin{document}

\title{Origin and Transport Signatures of Spin-Orbit Interactions in One- and Two-Dimensional SrTiO$_3$-Based Heterostructures}

\author{Younghyun Kim}
\affiliation{Department of Physics, University of California, Santa Barbara, California 93106}
\author{Roman M. Lutchyn}
\affiliation{Microsoft Research, Station Q, Elings Hall, University of California, Santa Barbara, California 93106}
\author{Chetan Nayak}
\affiliation{Microsoft Research, Station Q, Elings Hall, University of California, Santa Barbara, California 93106}
\affiliation{Department of Physics, University of California, Santa Barbara, California 93106}

\begin{abstract}
We study origin of Rashba spin-orbit interaction at SrTiO$_3$ surfaces and
LaAlO$_3$/SrTiO$_3$ interfaces by considering the interplay between atomic
spin-orbit coupling and inversion asymmetry at the surface or interface. We
show that, in a simple tight-binding model involving $3d$ $t_{2g}$ bands of
Ti ions, the induced spin-orbit coupling in the $d_{xz}$ and $d_{yz}$ bands
is cubic in momentum whereas the spin-orbit interaction in the $d_{xy}$ band
has linear momentum dependence. We also find that the spin-orbit interaction
in one-dimensional channels at LaAlO$_3$/SrTiO$_3$ interfaces is linear in
momentum for all bands. We discuss implications of our results for transport
experiments on SrTiO$_3$ surfaces and LaAlO$_3$/SrTiO$_3$ interfaces. In
particular, we analyze the effect of a given spin-orbit interaction term on
magnetotransport of LaAlO$_3$/SrTiO$_3$ by calculating weak
anti-localization corrections to the conductance and to universal conductance fluctuations.
\end{abstract}

\date{\today}
\maketitle

\section{Introduction}

The metallic interfaces \cite{Ohtomo04} between the
insulators LaAlO$_3$ (LAO) and SrTiO$_3$
(STO) exhibit both superconductivity \cite{Reyren07} and magnetism \cite{Brinkman07,Ariando11,Li11,Bert11} and, for some carrier densities,
they occur simultaneously \cite{Li11,Bert11,MannhartSchlom}.
Moreover, there is evidence that there is significant
spin-orbit interaction (SOI), as well \cite{Caviglia10,BenShalom,Triscone12}.
If all three of these phenomena are present,
then the ingredients are in place for topological superconductivity that
could support Majorana zero modes in confined structures
\cite{Read00,Kitaev01,Sau09,Lutchyn10,Oreg10,Alicea_PRB10}. Furthermore,
strong SOI could pave the way towards spintronics \cite{Zutic04} applications
of devices based on oxide interfaces.
In this paper, we give a simple microscopic understanding of the SOI
at LAO/STO interfaces induced by the combined effects of
atomic spin-orbit coupling and the interfacial electric field.

Caviglia {\it et al.} \cite{Caviglia10,Triscone12} and Ben Shalom {\it et al.} \cite{BenShalom}
made important experimental progress towards
understanding SOI effects at LAO/STO interfaces.
They found evidence that the magnetoconductance of LAO/STO
interfaces could be interpreted as resulting from weak anti-localization (WAL).
The SOI that they deduced showed strong dependence
on gate voltage, peaking at or near the gate voltage at which the
superconducting $T_c$ is maximized.
Nakamura {\it et al.} \cite{Nakamura'12} measured magnetoconductance
at the surface of STO and found that it could be fitted to a
cubic Rashba SOI. Zhong {\it et al.} \cite{Zhong12} performed a density functional theory (DFT) calculation, from
which they derived an effective tight-binding Hamiltonian. A key ingredient
supplied by the DFT calculation is the magnitude of
inter-orbital hopping terms.
They used the resulting tight-binding Hamiltonian to deduce a
Rashba-type energy splitting between the two spin components
of $\approx 2\, \text{meV}$ in the $d_{xy}$ band,
and a much larger splitting $\approx 20\, \text{meV}$ at the crossing point
of the $d_{xy}$ and $d_{xz}$ bands. Khalsa {\it et al.} \cite{Khalsa13}
further elucidated this by showing that inter-orbital hopping terms
are due primarily to the polar lattice displacement at the interface.

In this paper, we give a simple analysis of the Rashba SOI
in an effective tight-binding Hamiltonian for the $t_{2g}$ bands
of STO surfaces and LAO/STO interfaces, in order to better understand its basic qualitative features. Within our effective model, we find that Rashba SOI  is linear in momentum in  the $d_{xy}$ band, in agreement with Refs. \onlinecite{Zhong12,Khalsa13}, and is cubic in momentum in $d_{xz}$ and $d_{yz}$ bands.

We also consider one-dimensional(1D) channels at the LAO/STO interface.
Cen {\it et al.} \cite{Cen08} have fabricated such channels by `drawing'
them with an atomic force microscope (AFM) tip. Superconductivity is observed in these 1D channels\cite{Cen10,Veazey12}. Fidkowski
{\it et al.} \cite{Fidkowski13} proposed a theory for magnetism
and superconductivity in such channels in which conduction electrons in
the channels interact with localized spins to catalyze magnetic order
and interact with local superconducting fluctuations in STO to stabilize
quasi-long-ranged superconducting order.(For other theoretical perspectives, see Ref. \onlinecite{Marel11,Michaeli12,Fernandes13}.) In the presence of strong SOI,
such superconductivity can support Majorana zero modes at the ends of wires.
Conduction in such a channel will be dominated
by mobile $d_{xz}$ or $d_{yz}$ electrons, depending on the direction
of the channel. We show the SOI will be of linear-in-$k$ Rashba form in this case due to the broken rotational symmetry in the plane.

Based on the aforementioned conclusions regarding the different forms of the Rashba SOI in various geometries, we compute WAL correction to the magnetoconductance and show how one can distinguish different forms of Rashba SOI in transport experiments. We discuss the relevance of these results to understanding experiments at STO surfaces and LAO/STO interfaces. In the latter case, the
WAL signal can distinguish between transport dominated by the
$d_{xy}$ band; the $d_{xz,yz}$ bands; or 1D channels.
The 1D case should apply to channels `drawn' with an AFM tip
as well as to the 1D channels that appear to occur in putatively
two-dimensional(2D) systems~\cite{Kalisky2013} although, in the latter case, it will also
be important to account for the coupling between different 1D channels.

The paper is organized as follows. In Sec.~\ref{sec:model}, we introduce our effective tight binding model. The calculation of the effective spin-orbit interaction based on this model is presented in Sec. ~\ref{sec:SOI}.  In Secs.~\ref{sec:transport} and \ref{sec:exp}, we calculate weak anti-localization corrections to the conductance and discuss the manifestations of SOI in the context of recent experiments\cite{Caviglia10,BenShalom,Triscone12}.


\section{Three-Band Model}\label{sec:model}

\begin{figure}[tb]
\includegraphics[width=8cm]{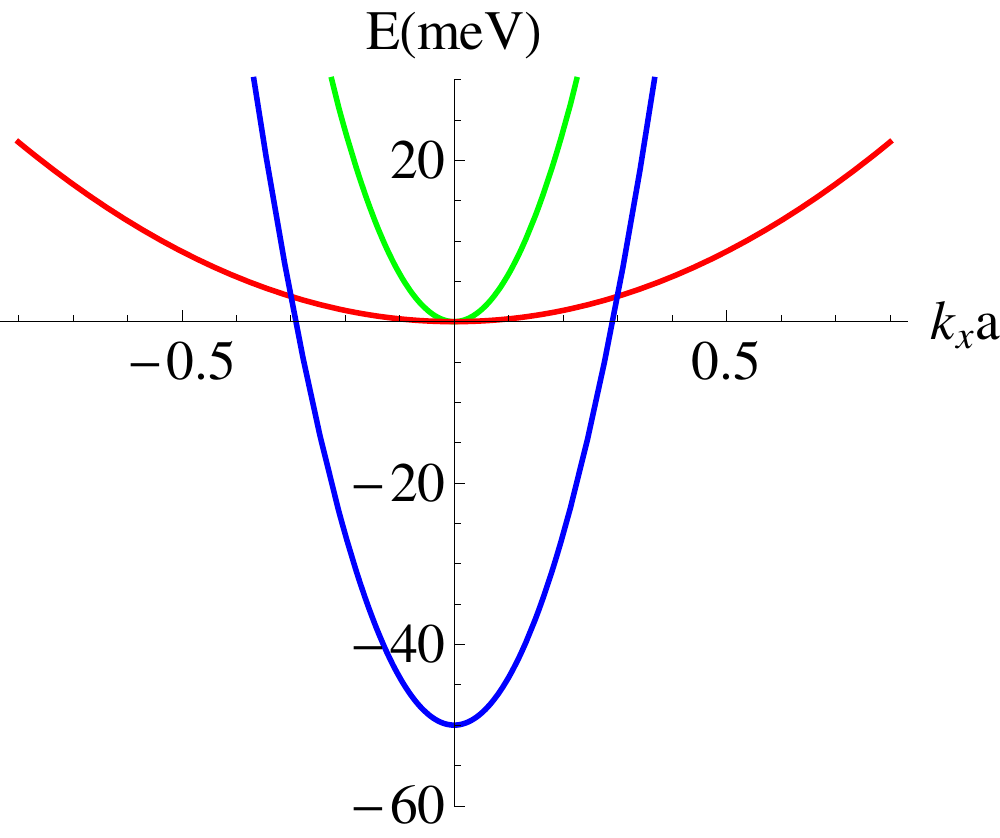}
\caption{Dispersion of $H_0$ for $\Delta_E=50 \,$meV \label{fig:h0band}}
\end{figure}

The Fermi energy at STO surfaces or STO-based interfaces
lies in the 3$d$ $t_{2g}$ bands of Ti ions near the surface/interface.
The $t_{2g}$ bands at the surface/interface are confined in $z$-direction,
which is normal to the surface/interface; consequently, they form
a two-dimensional electron gas (2DEG).
Here we consider only the top layer of STO and its $t_{2g}$ bands.
The Hamiltonian for these bands takes the form\cite{Bistritzer11,Joshua12}
\begin{align}\label{eq:Hamiltonian}
H=H_0+H_{ASO}+H_a
\end{align}
Here, $H_0$ is comprised of nearest-neighbor hopping and on-site interaction terms
that are diagonal in orbital space. In momentum space, it can be written in the form
${H_0} = {h_0}\otimes\sigma^0$, where $\sigma^0$ is the identity matrix
in spin space, and
\begin{align}
\!h_0\!=\!\left(\begin{array}{ccc}
\frac{\hbar^2k_x^2}{2m_h}\!+\!\frac{\hbar^2k_y^2}{2m_l}&0&0\\
0&\frac{\hbar^2k_x^2}{2m_l}\!+\!\frac{\hbar^2k_y^2}{2m_h}&0\\
0&0&\frac{\hbar^2k_x^2}{2m_l}\!+\!\frac{\hbar^2k_y^2}{2m_l}\!-\!\Delta_E\\
\end{array}\right)
\end{align}
Here $\Delta_E$ is the energy difference between the $d_{xy}$ band
and the $d_{xz},d_{yz}$ orbitals due to the confinement along $\hat z$-direction.
A recent DFT calculation\cite{Zhong12} suggests $m_l=0.41m_e$ and
$m_h=6.8m_e$ for bulk STO.
The second term in Eq.\eqref{eq:Hamiltonian} corresponds to the atomic spin-orbit coupling term which can be written in the form:
\ba
H_{ASO}&=&\frac{Zg\mu_B e}{16m_e c^2r^3\pi\epsilon_0} \,\vec{L}\cdot\vec{\sigma}\nonumber\\
&=&V_{ASO}\frac{Za_0^3}{\hbar}\frac{\vec{L}\cdot\vec{\sigma}}{r^3}
\ea
where the dimensionful prefactor $V_{ASO}=\frac{g\mu_B \hbar e}{16m_ec^2\pi\epsilon_0a_0^3}$, $\vec{L}=\vec{r}\times\vec{p}$ and $\vec{\sigma}=2\vec{S}/\hbar$ with $Z$ being the effective nuclear charge on the Ti atoms. The effective nuclear charge for the $d$-orbital electron in a neutral Ti atom $Z\approx 8.1$.

Atomic spin-orbit coupling projected to $t_{2g}$ orbital bands can be treated as an on-site orbital mixing term. Indeed, let's consider the limit ${m_l}, {m_h}\rightarrow\infty$ and compute matrix elements of the Hamiltonian $H_{ASO}$ between different orbital states:
\begin{multline}
\bra{j,d_{xz},\sigma'}H_{ASO}\ket{j,d_{xy},\sigma}=\\
V_{ASO}\frac{Za_0^3}{\hbar}\bra{j,d_{xz},\sigma'}\,\frac{\vec{L}\cdot\vec{\sigma}}{r^3}\ket{j,d_{xy},\sigma}
\end{multline}
where $\ket{j,d_{xy},\sigma}$ represents a state of an electron of spin $\sigma$
in the $d_{xy}$ orbital on site $r_j$. Given that $d_{xz}$ and $d_{xy}$ orbital wavefunctions are both odd in $x$, the matrix elements vanish by symmetry:
$\bra{j,d_{xz},\sigma'}{L_y}\ket{j,d_{xy},\sigma}=
\bra{j,d_{xz},\sigma'}{L_z}\ket{j,d_{xy},\sigma}=0$. The non-zero matrix element involves $d_{xz}$ and $d_{xy}$ bands
\begin{align}
\bra{j,d_{xz},\lambda'}&H_{ASO}\ket{j,d_{xy},\lambda}\nonumber\\
=& V_{ASO}\frac{Za_0^3}{\hbar}\bra{j,d_{xz},\lambda'}\,\frac{ L_x\sigma_x}{r^3}\ket{j,d_{xy},\lambda}\nonumber\\
=& V_{ASO}[\sigma_x]_{\lambda', \lambda}\frac{Za_0^3}{\hbar}\bra{j,d_{xz}}\,\frac{yp_z-zp_y}{r^3}\,\ket{j,d_{xy}}\nonumber\\
=& i\Delta_{ASO}\, [\sigma_x]_{\lambda', \lambda}.
\end{align}
In the last line, we have introduced the energy $\Delta_{ASO}$:
\ba
\Delta_{ASO}&=&V_{ASO}\,f(Z)
\ea
where the dimensionless form factor $f(Z)$ is defined as
\begin{align}
&f(Z)=\frac{Za_0^3}{i\hbar}\bra{j,d_{xz}}\,\frac{yp_z-zp_y}{r^3}\,\ket{j,d_{xy}}\nonumber\\
&=-\frac{2\,Z^8}{81^2\pi}\int_{-\infty}^\infty dx\,dy\,dz\,\frac{xz\,e^{-\frac{Zr}{3}}}{r^3}\left(y\frac{\partial}{\partial z}-\frac{\partial}{\partial y}z\right)xy\,e^{-\frac{Zr}{3}}\nonumber\\
&=\frac{Z^4}{405}.
\end{align}
%
%
%

\begin{figure}[h]
\includegraphics[width=8cm]{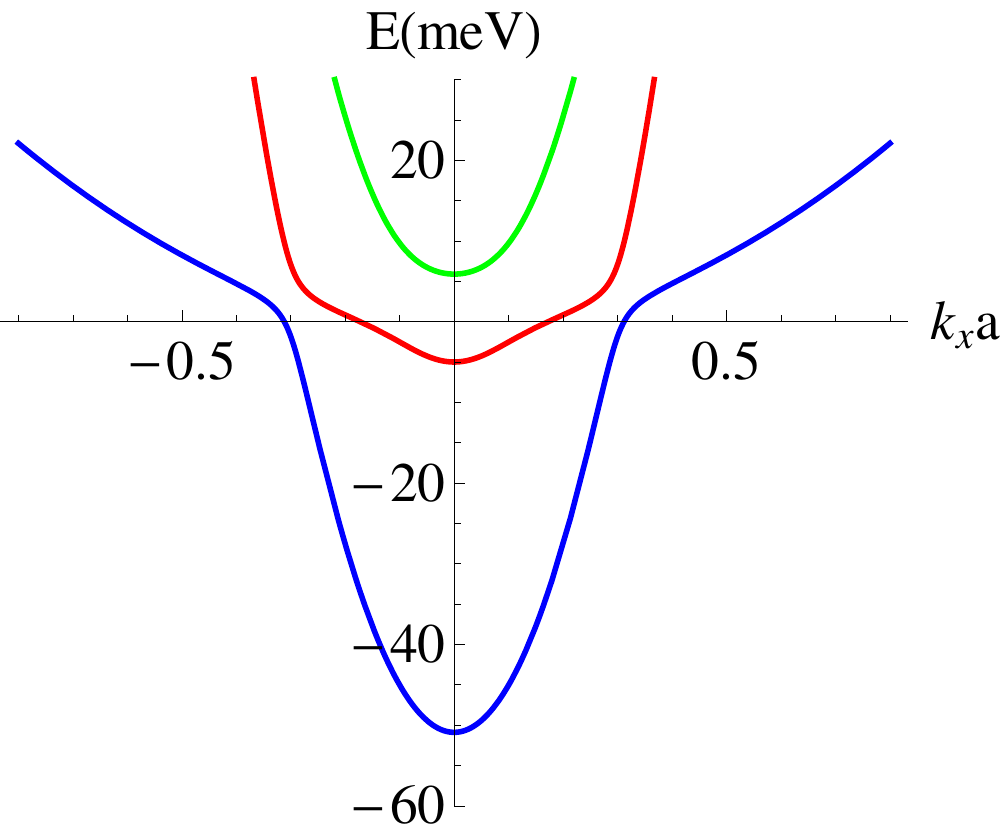}
\caption{Dispersion of $H_0+H_{ASO}$ with $\Delta_E=50 \,$meV and $\Delta_{ASO}=5 \,$meV \label{fig:hsoband}}
\end{figure}

Taking the matrix elements of $H_{ASO}$ between all three $t_{2g}$ orbitals
in a similar manner, an effective Hamiltonian in these bands,
$H^{t_{2g}}_{ASO}$ can be written as
\be
H^{t_{2g}}_{ASO}=\Delta_{ASO}\left(\begin{array}{ccc}
0&i\sigma_z&-i\sigma_y\\
-i\sigma_z&0&i\sigma_x\\
i\sigma_y&-i\sigma_x&0\\
\end{array}\right)
\ee
Fig. \ref{fig:hsoband} shows the non-degenerate band structure of $H_0+H_{ASO}$ for $\Delta_{ASO}=5 \,$meV. From the above
Hamiltonian, it may be seen that the lowest energy states and
highest energy states mix all three $t_{2g}$
orbitals (with selected spins), but the middle states only
contain $d_{yz}$ and $d_{xz}$ with same spin.

We now turn to inter-orbital nearest-neighbor hopping $H_a$, which induced primarily by polar lattice distortion due to the external electric from inversion asymmetry. We can qualitatively understand this as Ti-O-Ti hopping process between two neighbor Ti orbitals with different parity in $z$, for example, hopping between $d_{xy}-p_x-d_{xz}$ along y direction. Therefore, the effective form of $H_a$ in the basis of $t_{2g}$ orbital bands can be written as\cite{Zhong12,Khalsa13}
\be
H_a=\Delta_z\left(\begin{array}{ccc}
0&0&ik_x\\
0&0&ik_y\\
-ik_x&-ik_y&0\\
\end{array}\right)\otimes\sigma^0.
\ee

\begin{figure}[h]
\includegraphics[width=8cm]{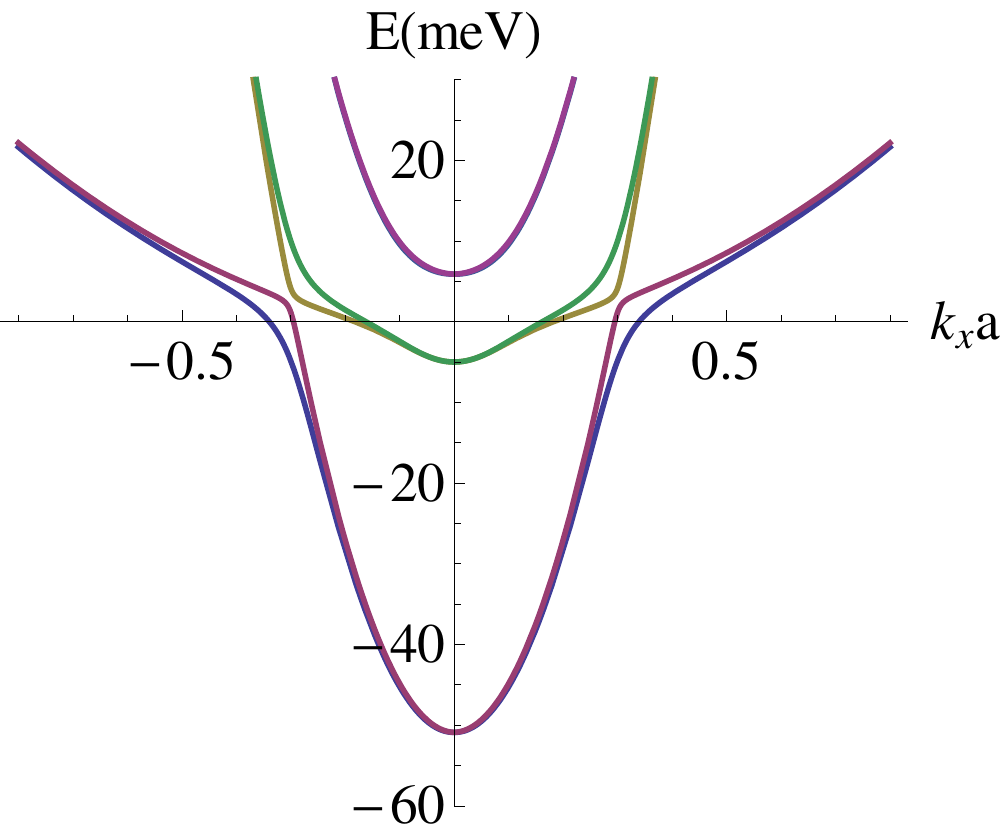}
\caption{Band structure corresponding to the Hamiltonian $H_0+H_{ASO}+H_a$ with $\Delta_E=50\,$meV, $\Delta_z=10\,$meV and $\Delta_{ASO}=5\,$meV. \label{fig:hrband}}
\end{figure}

Notice that $H_a$ generates hopping terms from $d_{xy}$ to $d_{xz}$ only in the
$y$-direction and from $d_{xy}$ to $d_{yz}$ only in the $x$-direction.
Otherwise, the hopping matrix element will be an integral of an odd function in
$x($or $y)$ and will vanish. Fig. \ref{fig:hrband} shows the spin splitted band structure of $H_0+H_{ASO}+H_a$ for $\Delta_{z}=10\,$meV and $\Delta_{ASO}=5\,$meV.


\section{Effective Spin-Orbit Interaction}\label{sec:SOI}

\subsection{Three-Band Model Near $k\sim0$}
For simplicity, we neglect the $k^2$ term in the energy dispersion in comparison to linear-in-$k$
terms. In this limit, we find the eigenstates of $H_0(\vec{k}=0)+H_{ASO}$, and then
express $H_a$ in that basis. The result is an effective Rashba SOI,
which takes the form:
\be
\frac{H_R}{\Delta_za} = \left(\begin{array}{cccccc}
	0	&	-\beta_1ik_-	&	0	&	\beta_2k_-	&	0	&	-\beta_3ik_+\\

	\beta_1ik_+	&	0	&	-\beta_2k_+	&	0	&	\beta_3ik_-	&	0	\\

	0	&	-\beta_2k_-	&	0	&	0	&	0	&	-\beta_4k_-	\\

	\beta_2k_+	&	0	&	0	&	0	&	\beta_4k_+	&	0	\\

	0	&	-\beta_3ik_+	&	0	&	\beta_4k_-	&	0	&	\beta_1ik_+	\\

	\beta_3ik_-	&	0	&	-\beta_4k_-	&	0	&	-\beta_1ik_-	&	0	\\

\end{array}\right)
\label{eqn:erh}
\ee
Here $k_\pm=k_x\pm ik_y$ and the order of $H_0(\vec{k}=0)+H_{ASO}$ eigenstates is from highest energy to lowest energy. That means $(1,0,0,0,0,0)$ and $(0,1,0,0,0,0)$ correspond to the highest energy eigenstates of $H_0(\vec{k}=0)+H_{ASO}$. Note that there are three energy
eigenvalues for $H_0(\vec{k}=0)+H_{ASO}$, with two
Kramers-degenerate eigenstates for each.
From this Hamiltonian we expect linear in momentum Rashba SOI (linear Rashba SOI) in bottom and top bands and cubic in momentum SOI(cubic Rashba SOI) $(\Delta\sim\alpha_3k^3)$ in the middle band. The absence of linear Rashba SOI in the middle band is due to the fact that the middle band at $k\sim0$ contains only $d_{xz}$ and $d_{yz}$ components and, therefore, is odd in $z$-direction.

\begin{figure}[h]
\includegraphics[width=8cm]{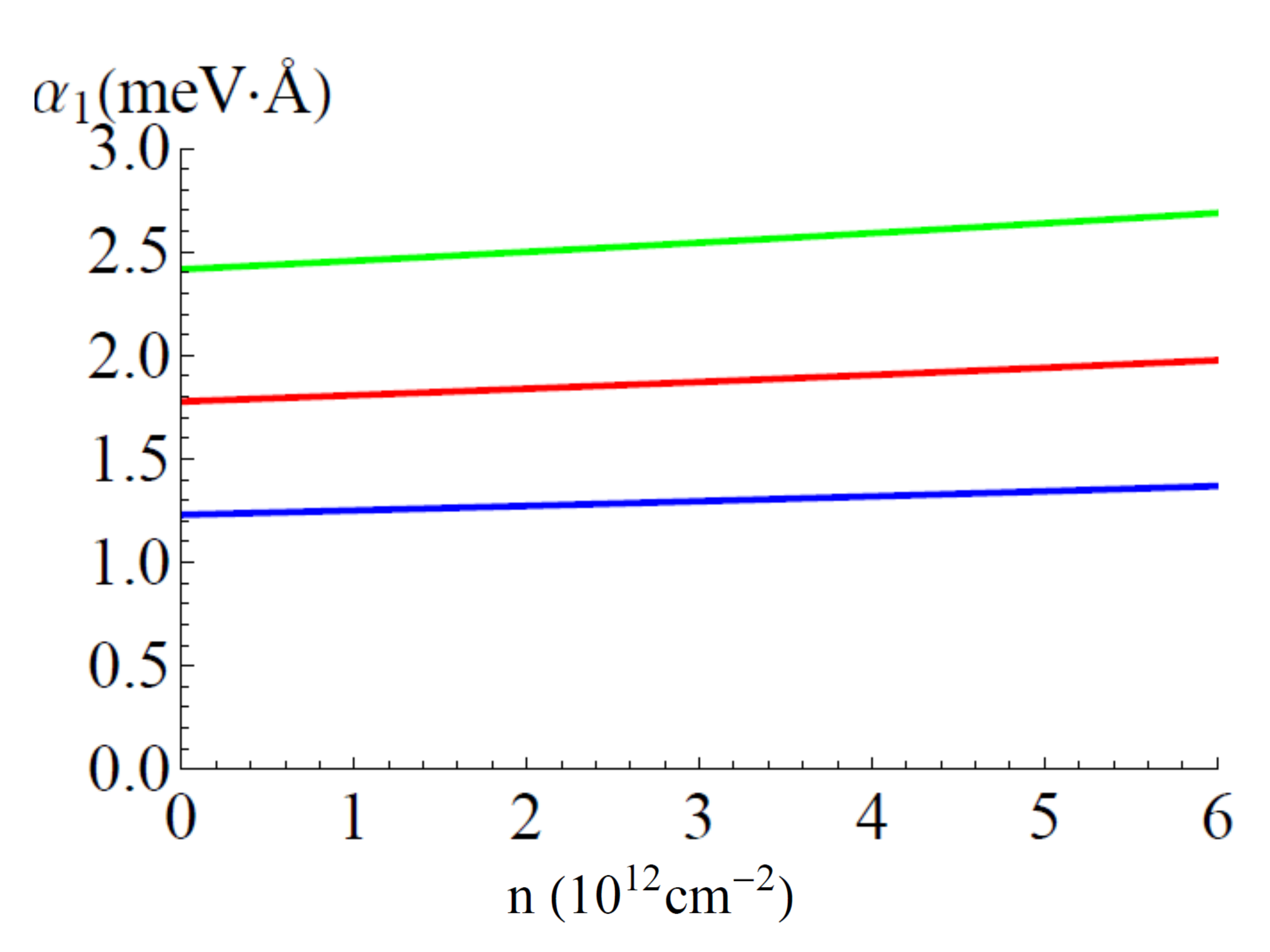}
\caption{$\alpha_1\,$(meV$\cdot\AA$) vs $n\,$($10^{12}$cm$^{-2}$) at STO surfaces. Upper curve: $\Delta_z=10\,$meV, $\Delta_{ASO}=5\,$meV. Middle curve: $\Delta_z=5\,$meV, $\Delta_{ASO}=15\,$meV. Bottom Curve: $\Delta_z=10\,$meV, $\Delta_{ASO}=10\,$meV. \label{fig:a1STO}}
\end{figure}

The coupling coefficients $\beta_i$ depend on $\Delta_{ASO}$ and $\Delta_E$. When the band splitting $\Delta_E$ is much larger than $\Delta_{ASO}$, the lowest band is primarily $d_{xy}$-like near $k\sim0$, and we can estimate the size of $k-$linear Rashba coupling in the lowest energy bands $\alpha_1=\Delta_z \,a\, \beta_1$ from the second-order perturbation, i.e. first-order in $H_{ASO}$ and first-order in $H_a$ in orbital basis as follows:
\begin{multline}
\bra{k,d_{xy},\sigma'}H^{(2)}\ket{k,d_{xy},\sigma}=\\
\sum_{k',\sigma''}\frac{\bra{k,d_{xy},\sigma'}H_a\ket{k',d_{xz},\sigma''}\bra{k',d_{xz},\sigma''}H_{ASO}\ket{k,d_{xy},\sigma}}{E_{d_{xz}}(k')-E_{d_{xy}}(k)} \\
+ (d_{xz}\rightarrow d_{yz})\\
=\frac{\Delta_{ASO}\Delta_z}{\Delta_{BG}(k)}(\left\langle\sigma_y\right\rangle
\sin(k_xa)-\left\langle\sigma_x\right\rangle\sin(k_ya))\\
\sim \alpha_1 (\vec{k}\times\vec{\sigma})\cdot\hat{z}
\end{multline}
where
\ba
\Delta_{BG}(k)&=&E_{d_{xz}}(k)-E_{d_{xy}}(k)\sim\Delta_E\\
\alpha_1&\sim&\frac{\Delta_{ASO}\Delta_z}{\Delta_E}a
\ea
This perturbative description of the Rashba SOI breaks down at the band crossings of $H_0$, and we find dramatic changes in the strength of SOI as we will see in Sec IV. with exact diagonalization analysis. However, we find linear Rashba SOI dominates cubic Rashba SOI for small $k$. To see this in more detail, we restore the $k^2$ energy dispersion in Eq. (\ref{eqn:erh}) and compute $H_R$ retaining both linear and cubic terms. We take $\Delta_E=320\,$meV\cite{Zhong12} and plot the strength of SOI $\alpha_1$ as a function of carrier density for several values for $\Delta_{ASO}$ and $\Delta_z$ as we can see in Fig. \ref{fig:a1STO}. Here $\alpha_1=\Delta_R/k_x$ where $\Delta_R$ is the Rashba SOI-induced energy splitting of the bottom bands.
 The linear Rashba coupling can be identified as the value of $\alpha_1$ at $k=0$. The slope of the plot is proportional to the cubic Rashba effect. For various values for $\Delta_{ASO}$ and $\Delta_E$, we see that the contribution from the cubic term is dominated by the linear term.

\subsection{Four-Band Model}
\begin{figure}[h]
\includegraphics[width=8cm]{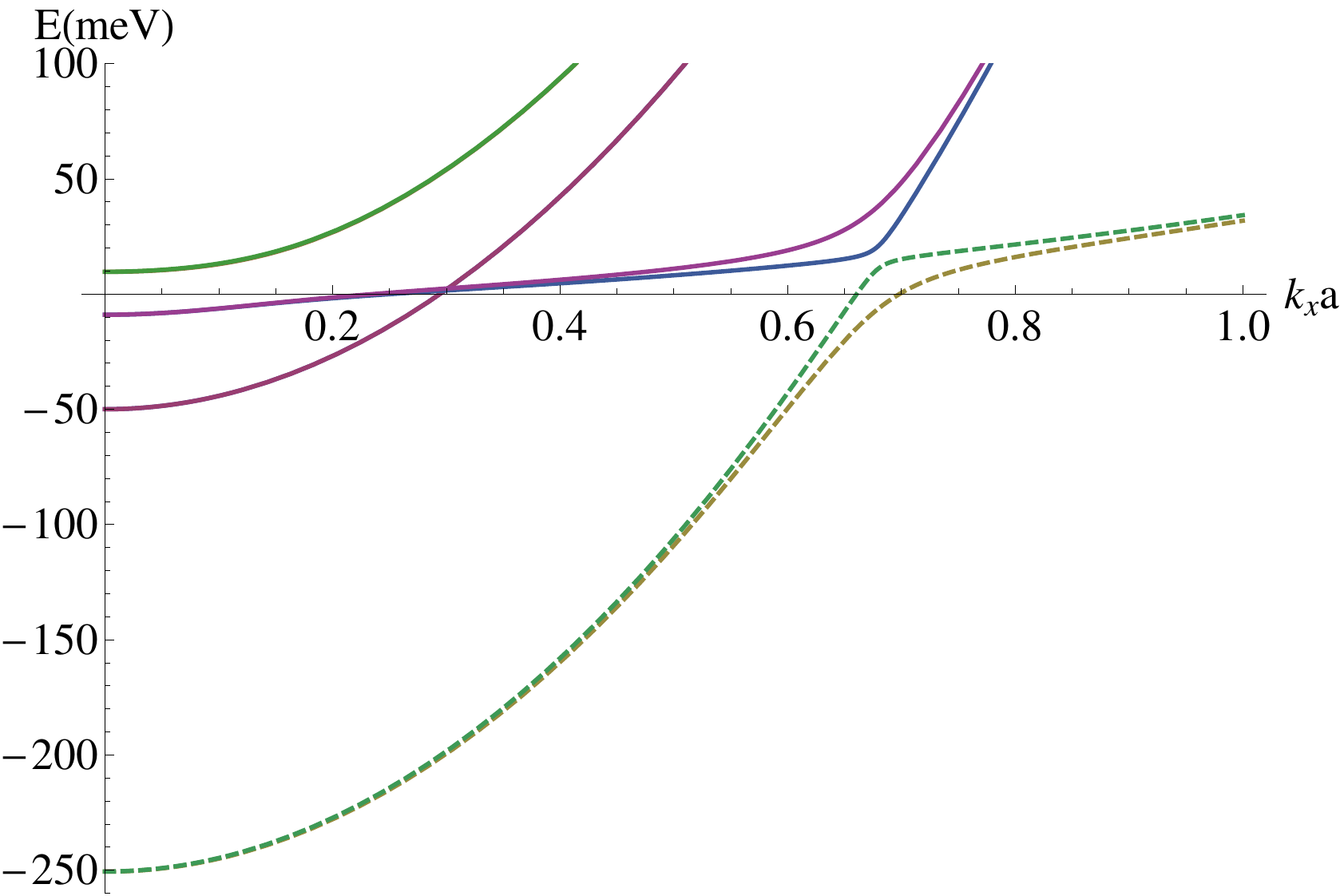}
\caption{Band structure of 4-band model with $\Delta_{ASO}=9\,$meV and $\Delta_z=20\,$meV. The first $d_{xy}$ sub-band(dashed lines) is assumed to be localized. \label{fig:b4LAOSTO}}
\end{figure}
We can extend our theory to four-band model in which two $d_{xy}$ sub-bands lies below $d_{xz}$ and $d_{yz}$ band due to the strong confinement along $z$-axis(See Fig. \ref{fig:b4LAOSTO}). Recent DFT calculation\cite{Zhong12} shows the first (second) $d_{xy}$ sub-band has $\Delta_{E}=250(50)\,$meV for LAO/STO interface. Most of the electrons coming from the polar catastrophe$\sim 10^{14}\,$cm$^{-2}$ are localized in the first $d_{xy}$ sub-band as suggested in density functional calculation of Ref. \onlinecite{Pentcheva06,Popovic08}. The main difference with the three-band model is that the second $d_{xy}$ sub-band has a much smaller $\Delta_z$ since they do not see the large electric field that the first sub-band electrons see. Hence, those light electrons do not contribute to the anti-localization effect in our picture.

\subsection{Effective Model for Quasi-One-Dimensional Channel}
Now we will consider a quasi one-dimensional system at the LAO/STO interface which can be related to a nanowire artificially drawn using AFM tip with LAO(3.u.c.)/STO interface\cite{Cen08,Levy1}. We assume there is a confinement in $y$-direction such that wavevector in $y$-direction is quantized as $k_y^n=\frac{\pi n}{w}$, and degeneracy between $d_{xz}$ and $d_{yz}$ bands at $k=0$ is lifted. With quantized $k_y$, dispersion relations can be written as,
\ba
E^{xz}(k_x)&=&\frac{\hbar^2k_x^2}{2m_l}-\Delta_{Ey}\nonumber\\
E^{yz}(k_x)&=&\frac{\hbar^2k_x^2}{2m_h}\\
E^{xy}(k_x)&=&\frac{\hbar^2k_x^2}{2m_l}-\Delta_{Ez}\nonumber
\ea
where $\Delta_{Ey}(\Delta_{Ez})$ is the energy splitting due to the $y(z)$-direction confinement. Since the degeneracy is lifted, the $d_{xz}$-like band also has $k-$linear spin-orbit coupling, and it seems hard to distinguish it from $d_{xy}$-like band with weak anti-localization measurement. However, we find that the relation between chemical potential and spin-orbit coupling strength strongly depends on the band.
\begin{figure}[h]
\includegraphics[width=8cm]{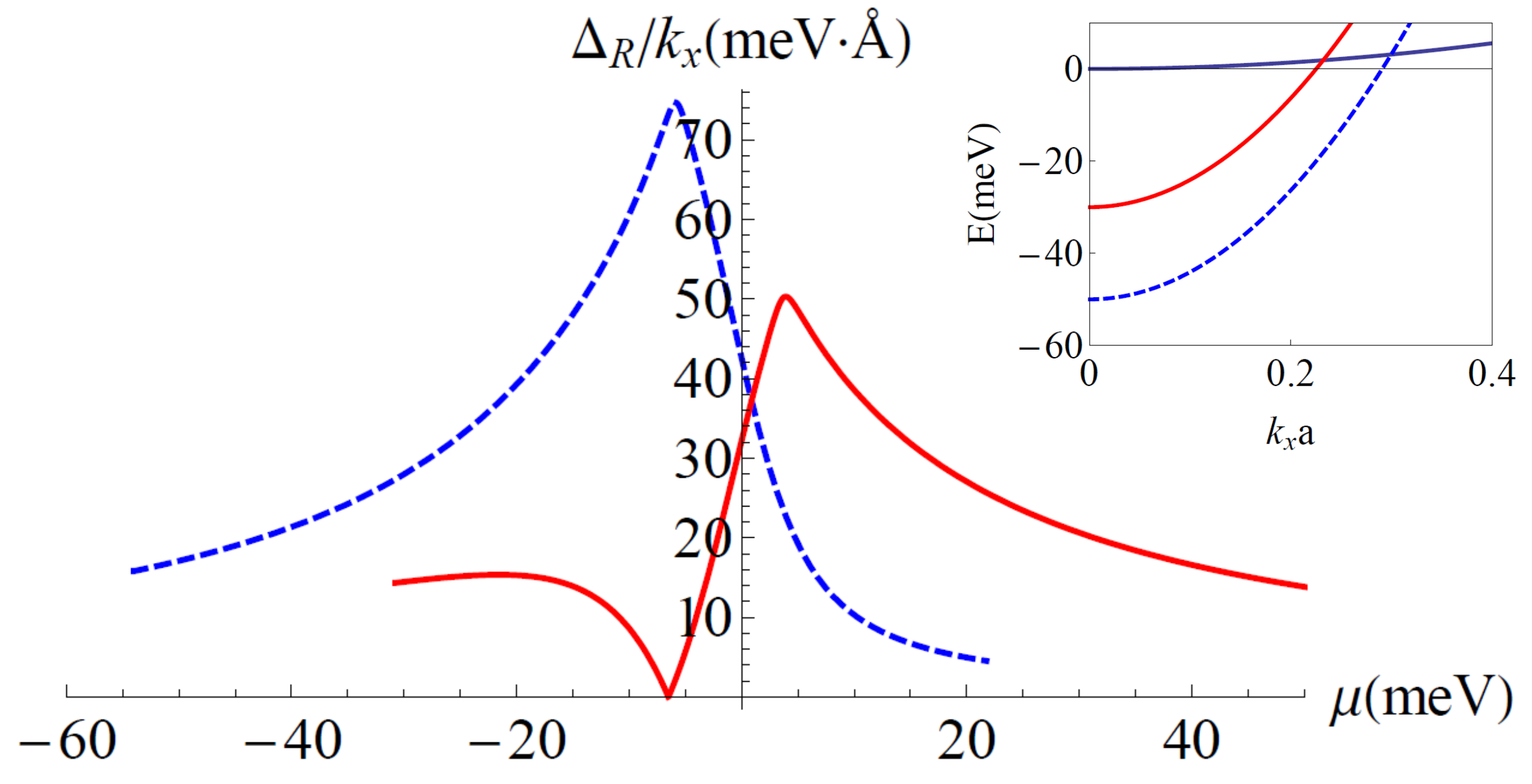}
\caption{SOI strength versus chemical potential for quasi one-dimensional system. The inset shows dispersion of $H_0$. Dashed(Solid) curve is $d_{xy}(d_{xz})$-like band. $\Delta_{Ey}=30\,$meV, $\Delta_{Ez}=50\,$meV, $\Delta_{ASO}=9\,$meV and $\Delta_z=20\,$meV are used. SOI of $d_{xz}$-like band changes sign at crossing point of $d_{xz}$ and $d_{yz}$ band(two solid lines in inset).
\label{fig:1d}}
\end{figure}
As may be seen in Fig.\ref{fig:1d}, if the transport is still dominated by $d_{xz}$-like band, we can see that strength of SOI goes to zero at specific value of $\mu_0 ($or $k_0)$ because the degeneracy between $d_{xz}$ and $d_{yz}$ band is recovered at the band crossing point of $H_0$. When the transport is dominated by $d_{xy}$-like band, the evolution of SOI strength as increasing chemical potential does not have any nodes.


\section{Effect of spin orbit interaction on magnetoconductivity}\label{sec:transport}
\subsection{Quantum Corrections to Conductivity in Two Dimensions}

The presence of significant SOI changes the universality class of the Hamiltonian, and results in a dramatic difference in weak field magnetoconductance predictions\cite{Hikami80}. Indeed, it is well known that SOI leads to a sign change of the quantum correction to conductivity $\Delta \sigma $. This phenomenon, known as WAL,
can be used a diagnostic for the presence of SOI in a conductor.
In principle, modification of band structure by SOI
can be observed in Shubnikov-de Haas oscillations, but this effect
will be washed out if impurity scattering is too large. The frequency
difference between the two Fermi surfaces, $\Delta \omega$
must satisfy $\Delta \omega \,\tau \gg 1$.
On the other hand, WAL is observable under the
much less stringent condition, $\sigma_{xx} \gg \frac{e^2}{h}$.
From a detailed fit of the dependence of WAL as a function of the density,
it is possible to deduce the form of SOI, i.e. in our context whether SOI is linear or cubic in momentum
and, in the former case, whether it is due to 1D or 2D transport.

In the subsections that follow, by treating the magnetic field as
a long-distance cutoff, we derive (relatively) simple
closed forms of the WAL corrections to the conductivity in the limits of (a) purely linear
2D Rashba SOI; (b) purely cubic 2D Rashba SOI; and (c)
linear quasi-1D Rashba SOI. In the first two regimes and in the limit
of small magnetic fields, we recover the Iordanskii, Lyanda-Geller, Pikus (ILP) theory
\cite{Iordanskii94}, which treats the magnetic field more
precisely by summing over Landau levels. In the third regime,
we obtain similar results to those of Kettemann \cite{Kettemann07}.

We argue that it is essential to use these forms to fit WAL
data for STO surfaces and LAO/STO interfaces.
The ILP theory was used by
Nakamura {\it et al.}\cite{Nakamura'12}
to deduce the cubic $k$-dependence of Rashba SOI
at the surface of STO. (See Eq. (S1) of Ref. \onlinecite{Nakamura'12}.)
However, Caviglia {\it et al.} \cite{Caviglia10} used the
Maekawa-Fukuyama (MF) theory\cite{Maekawa81}, which
incorporates the SOI simply as a spin relaxation time, following
Hikami, Larkin, Nagaoka (HLN) theory\cite{Hikami80}.
In the limit of weak SOI and weak Zeeman splitting.
the MF theory gives similar results to the ILP theory
with dominant SOI given by cubic Rashba.
Therefore, Caviglia et al.'s analysis\cite{Caviglia10} could be
understood as an indication that their
results fit the ILP theory with cubic Rashba, as at the surface of STO.

However, since the MF theory and the HLN theory, on which it was based,
were clearly formulated for a physically-distinct situation that is
not applicable to the LAO/STO interface, it is necessary to compare magnetoconductance data to an appropriate theory that takes as its starting point either linear or cubic Rashba SOI (in 2D or 1D).
We perform such an analysis in the subsections that follow and suggest a method for fitting WAL at LAO/STO interfaces in the low carrier density region which lead us to distinguish the contribution from linear and cubic Rashba effects. We discuss experimental results from this perspective in the following section.


\subsubsection{Weak anti-localization due to linear Rashba spin-orbit interaction}

Assuming that the dominant contribution to current transport is coming from $d_{xy}$ band, we only need to consider $k$-linear Rashba SOI for small carrier density. Therefore, the effective Hamiltonian reads:
\be
H=\frac{\hbar^2k^2}{2m_l}+\hbar\,\pmb{\sigma} \cdot \pmb{\Omega}_{R1}
\ee
where $\pmb{\sigma}=(\sigma_x,\sigma_y)$ is a vector of Pauli spin matrices, $\pmb{\Omega}_{R1}=\Omega_{R1}(\sin\theta,-\cos\theta)$ is an effective magnetic field of linear Rashba SOI, $k=\sqrt{k_x^2+k_y^2}$, $\Omega_{R1}=\alpha_1 k/\hbar$ and $\tan\theta=k_x/k_y$. We now consider scattering on short-range impurities. For uncorrelated Gaussian disorder potential $V(x)$, characterized by the correlation function $\langle V(x)V(x')\rangle=\frac{1}{2\pi \nu \tau_0}\delta(x-x')$, quantum corrections to the dc conductivity are given by~\cite{Hikami80}
\begin{align}
\Delta \sigma=-\frac{2e^2}{h}D \sum_{\alpha, \beta} \int \frac{d^2q}{(2\pi)^2} 2\pi\nu\tau_0^2 C_{\alpha \beta \beta \alpha}(q).
\end{align}
Here $D$, $\nu$ and $\tau_0$ are the diffusion constant, 2D density of states and elastic mean-free time, respectively; $C_{\alpha \beta \beta \alpha}(q)$ is disorder-averaged Cooperon propagator with $\alpha, \beta$ being spin indices.  Following ILP's approach\cite{Iordanskii94}\cite{Knap96}, the matrix equation for the zeroth harmonic of the Cooperon propagator which gives the dominant contribution to WAL in the diffusive limit, reads
\be
\hat{\mathcal{L}}\hat{C}_0(q)=\frac{1}{2\pi\nu\tau_0^2}.
\label{eqn:ME}
\ee
It is convenient to rewrite Cooperon propagator in the angular momentum basis, in which the singlet $J=0$ and triplet $J=1$ sectors are decoupled. The eigenvalue for the singlet contribution can be readily obtained $E^0=D(q^2+q_\phi^2)$ where $q_\phi^2=1/D\tau_\phi$ with $\tau_\phi$ being the inelastic scattering time. Henceforth, we consider the triplet $J=1$ sector, and find the corresponding eigenvalues. The latter requires to diagonalize $\!\hat{\mathcal{L}}_{J=1}\!$
\begin{align}\label{linearR}
\!\!&\!\hat{\mathcal{L}}_{J=1}\!\!=\!\!Dq^2\!+\!\frac{1}{\tau_\phi}\!+\!2\Omega^2_{R1}\tau_1(\hat {J}^2\!-\!\hat{J}_z^2)\!+\!iv_F\tau_1\Omega_{R1}\!(\hat{J}_+q_-\!-\!\hat{J}_-q_+),\\
&\frac{1}{\tau_0}=\int W(\varphi)d\varphi, \,\,\,\,\,\,\,\, \frac{1}{\tau_n}=\int W(\varphi)(1-\cos(n\varphi))d\varphi,\nonumber\\
&\hat{J}_\pm=\hat{J}_x\pm i \hat{J}_y, \,\,\,\,\,\,\,\,\,\,\,\,\,\,\,\,\,\,\, q_\pm=q_x\pm i q_y.\nonumber
\end{align}
Here $D=v_F^2\tau_1/2$ is 2D diffusion constant, $W(\varphi)$ is scattering rate for an angle $\varphi$, and $\hat{J}_i$ are vector components of the total angular momentum operator. The SOI mixes different components of the $J=1$ manifold of the Cooperon propagator. By diagonalizing $\hat{\mathcal{L}}$, one finds
\begin{align}\label{eq:linearRashba}
\frac{E^0}{D}&=q^2+q_\phi^2\\
\frac{E^1_{0}}{D}&=q^2+q_\phi^2+q_{so}^2\nonumber\\
\frac{E^1_-}{D}&=q^2+q_\phi^2+\frac{3}{2}q_{so}^2-\sqrt{4q^2q_{so}^2+\frac{q_{so}^4}{4}}\nonumber\\
\frac{E^1_{+}}{D}&=q^2+q_\phi^2+\frac{3}{2}q_{so}^2+\sqrt{4q^2q_{so}^2+\frac{q_{so}^4}{4}},\nonumber
\end{align}
where $E^0$ and $E^1_m$ are eigenvalues of $\hat{\mathcal{L}}$ corresponding to total angular momentum $J=0$ and $J=1$ sectors, and  $q_{so}^2=2\Omega^2_{R1}\tau_1/D=2\alpha^2k_F^2\tau_1/\hbar^2D$ characterizes the strength of SOI. Using these results, one can obtain WAL correction to conductivity:
\begin{align}
\Delta\sigma&=-\frac{2e^2}{h}D\int_{q_{\rm min}}^{q_{\rm max}} \frac{d^2q}{(2\pi)^2} 2\pi  \nu\tau_0^2 {\rm Tr} [ \hat{C}(q)]\nonumber\\
&=-\frac{2e^2}{h}D\int_{q_{\rm min}}^{q_{\rm max}}\frac{d^2q}{(2\pi)^2}\left(-\frac{1}{E^0}+\sum_{m=-1}^{1}\frac{1}{E^1_m}\right).
\end{align}
Here $q_{\rm max}$ and $q_{\rm min}$ are ultra-violet and infra-red cutoffs, respectively, with $q_{\rm max} = 1/v_F\tau_1$. If magnetic field is weak, one can simplify the calculation by including magnetic field as IR cutoff given by $q_{\rm min}=q_B$. (If magnetic field is large, one has to perform the summation over the Landau levels\cite{Iordanskii94}). In this paper, we focus on weak magnetic field limit, in which case one can obtain analytical expression for the quantum correction to magnetoconductivity:
\begin{align}
\Delta\sigma(B)&-\Delta\sigma(0)=-\frac{e^2}{2\pi h}(-\Delta I^0+\Delta I^1_0+\Delta I^1_{-}+\Delta I^1_{+})\nonumber\\
\Delta I^0&=\ln\left[\frac{q_\phi^2}{q_\phi^2+q_B^2}\right], \,\,\, \Delta I^1_0=\ln\left[\frac{q_\phi^2+q_{so}^2}{q_\phi^2+q_{so}^2+q_B^2}\right],\nonumber\\
\Delta I^1_-&=\ln\left[\frac{q_\phi^2+q_{so}^2}{q_\phi^2+q_{so}^2+q_B^2}\right]+F_-(q_{so}, q_B, q_\phi),\nonumber\\
\Delta I^1_+&=\ln\left[\frac{q_\phi^2+2q_{so}^2}{q_\phi^2+2q_{so}^2+q_B^2}\right]+F_+(q_{so}, q_B, q_\phi)
\label{eqn:WAL}.
\end{align}
Here $q_B^2\sim eB/\hbar$, and the functions $F_-(q_{so}, q_B, q_\phi)$ and $F_+(q_{so}, q_B, q_\phi)$ are defined as
\begin{align}
&F_-(q_{so}, q_B, q_\phi)=-\frac{4\pi q_{so}}{q_F}\left[ 1  -
\frac{1}{\sqrt{1+16({q_B}/{q_F})^2}}  \right]\nonumber\\
     &-\frac{8q_{so}}{q_F}	\left[	\arctan\left( \frac{3q_{so}}{q_F}\right) -\frac{\arctan\left( \frac{3q_{so}}{\sqrt{q_F^2+16q_B^2}}\right)}{\sqrt{1+16({q_B}/{q_F})^2}}   	\right]\nonumber\\
&F_+(q_{so}, q_B, q_\phi)=\frac{4\pi q_{so}}{q_F}\left[1  -
\frac{1}{\sqrt{1+16({q_B}/{q_F})^2}} \right]\nonumber\\
     &-\frac{8q_{so}}{q_F}	\left[	\arctan\left( \frac{5q_{so}}{q_F}\right) -\frac{\arctan\left( \frac{5q_{so}}{\sqrt{q_F^2+16q_B^2}}\right)}{\sqrt{1+16({q_B}/{q_F})^2}}   	\right].
\nonumber
\end{align}
with $q_F^2 = 7q_{so}^2+16q_\phi^2$. The magnetic field cutoff $q_B^2$ depends on an arbitrary coefficient $a_1$ that enters the definition $q_B^2=a_1 eB/\hbar$. By comparing Eq.~\ref{eqn:WAL} with the ILP theory in the limit $q_\phi^2\ll q_B^2 \ll q_{so}^2$, we find that our results coincide if one takes $a_1=e^{-\gamma}$ with $\gamma$ being Euler's constant.


\subsubsection{Weak anti-localization due to cubic Rashba spin-orbit interaction}
According to our four-band model, there are two types of carriers contributing to the transport. One is the second level $d_{xy}$ band electrons with negligible SOI, which we therefore
neglect, and the other is middle ($d_{xz} + d_{yz}$) band electrons with $k$-cubic Rashba
SOI. In this case, the effective Hamiltonian for middle band derived from Eq. \ref{eqn:erh} can be written as
\be
H=\frac{\hbar^2k^2}{2m^*}+\hbar\,\pmb{\sigma} \cdot \pmb{\Omega}_{R3},
\ee
where $\pmb{\Omega}_{R3}=\Omega_{R3}(\sin3\theta,-\cos3\theta)$ with $\Omega_{R3}=\alpha_3k_F^3/\hbar$ now gives correction to third harmonic of Cooperon, and resulting matrix equation for Cooperon in the triplet sector can be written as Eq. \ref{eqn:ME} with
\be
\hat{\mathcal{L}}_{J=1}=Dq^2+\frac{1}{\tau_\phi}+2\Omega^2_{R3}\tau_3(J^2-J_z^2).\\
\ee
Now $\hat{\mathcal{L}}_{J=1}$ is readily diagonal in the original basis of $J$ and $J_z$, and its eigenvalues are
\begin{align}
\frac{E^0}{D}&=q^2+q_\phi^2,  \,\,\,\,\,\,\,\,\,\,\,\,\,\,\,\,\,\,\,\, \frac{E^1_{-1}}{D}=q^2+q_\phi^2+q_{so3}^2,\nonumber\\
\frac{E^1_0}{D}&=q^2+q_\phi^2+2q_{so3}^2, \,\,\,\, \frac{E^1_{1}}{D}=q^2+q_\phi^2+q_{so3}^2,\nonumber
\end{align}
where $q^2_{so3}=2\Omega^2_{R3}\tau_3/D=2\alpha_3^2k_F^6\tau_3/\hbar^2D$.
Then the magnetoconductivity can be written as,
\begin{align}
\Delta\sigma(B)-&\Delta\sigma(0)=-\frac{e^2}{2\pi h}(-\Delta I^0+\Delta I^1_0+\Delta I^1_{-}+\Delta I^1_{+})\nonumber\\
\Delta I^0=\ln&\left[\frac{q_\phi^2}{q_\phi^2+q_B^2}\right], \,\,\, \Delta I^1_0=\ln\left[\frac{q_\phi^2+2q_{so3}^2}{q_\phi^2+2q_{so3}^2+q_B^2}\right],\nonumber\\
\Delta I^1_-=\ln&\left[\frac{q_\phi^2+q_{so}^2}{q_\phi^2+q_{so3}^2+q_B^2}\right], \nonumber\\
\Delta I^1_+=\ln&\left[\frac{q_\phi^2+q_{so3}^2}{q_\phi^2+q_{so3}^2+q_B^2}\right]
\label{eqn:WAL-cubic}
\end{align}
Again, taking $q_B^2=e^{-\gamma} Be/\hbar$ reproduces HLN theory\cite{Hikami80} for strong SOI and small magnetic field.\\

Clearly, WAL corrections to the magnetoconductance are different for linear and cubic Rashba SOI. Therefore, by comparing fits of the experimental data to the above expressions, one can try to distinguish between the two scenarios. This, in turn, can shed light on the origin of superconductivity at the LAO/STO interface.

\subsection{Universal Conductance Fluctuations in Two Dimensions}
\begin{figure}[h]
\includegraphics[width=8cm]{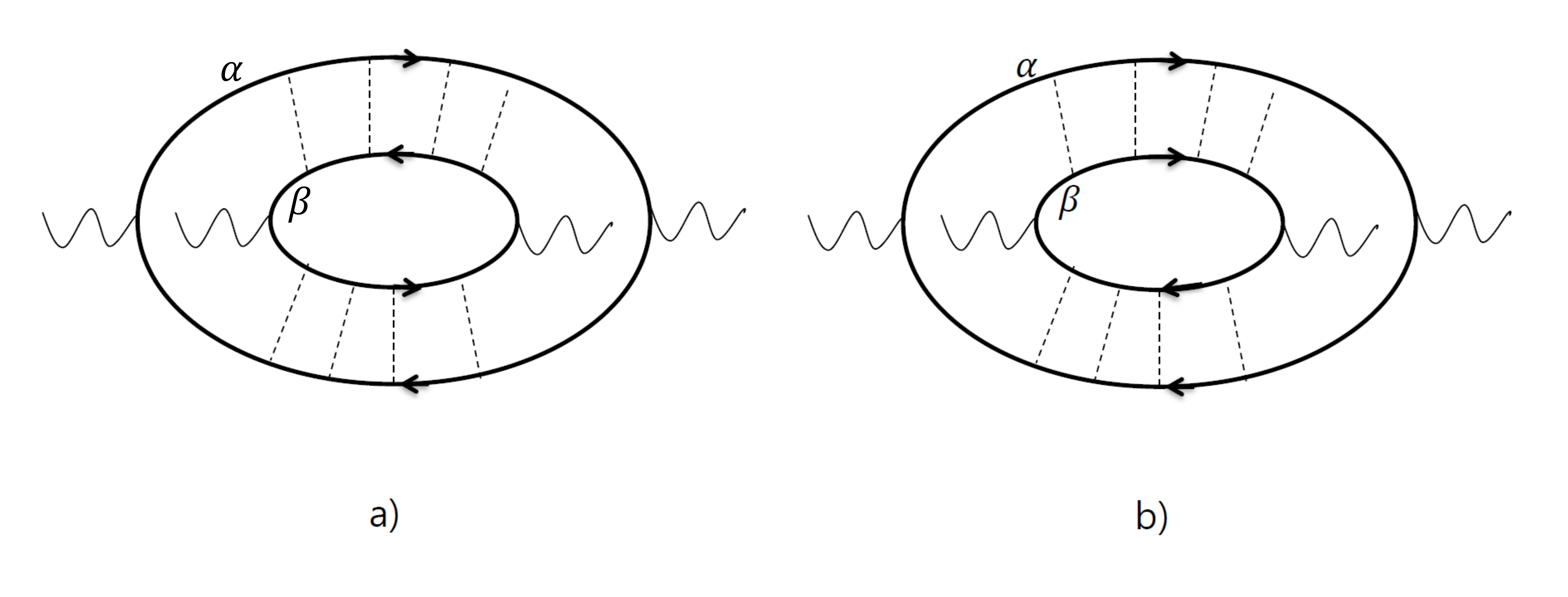}
\caption{a) Diagrams corresponding to two dominant contributions to UCF: a) particle-hole channel b) particle-particle channel.}
\label{fig:ucfdiagram}
\end{figure}

Another transport signature of the spin-orbit interaction is its effect on universal conductance fluctuations (UCF) in small systems (see Ref~.\onlinecite{Akkermans07}
and references therein).
The variance of the conductivity in a mesoscopic system, $\delta\sigma^2$, has dominant contributions from the two types of connected diagrams shown in Fig. \ref{fig:ucfdiagram}.
We assume that our system has size $L$ in each direction so that its area $V$ is given by $V=L^2$. The contributions from those diagrams can written in the form
\ba \label{eq:UCF}
\delta\sigma^2\sim \left(\frac{e^2}{h}\right)^2\cdot\frac{D^2}{V}\int^{q_{max}}_{q_{min}}\frac{d^2q}{(2\pi)^2}\left( {\rm Tr}[\hat{D}(q)]^2 + {\rm Tr} [\hat{C}(q)]^2\right)\nonumber\\
\ea
where $D_{\alpha \beta}(q)= \delta_{\alpha \beta} /D(q^2+q_{\rm IR}^2)$ is a Diffuson propagator, and $\hat{C}(q)$ is a Cooperon propagator derived in the previous section. We will assume $q_{\rm max}=1/l\rightarrow \infty$, $q_{\rm IR}=\text{max}[1/L,1/l_\phi]$; the cutoff $q_{\rm min}=0$ and $q_{\rm min}=q_B$ for Diffuson and Cooperon propagators, respectively.

\subsubsection{UCF for linear Rashba spin-orbit coupling}
Taking into account finite-size effects in Eq.~\eqref{linearR}, one finds Cooperon propagators for linear Rashba SOI
\ba
C^0_0&=&\frac{1}{D(q^2+q_{IR}^2)}, \,\,\,\,\, C^1_0=\frac{1}{D(q^2+q_{IR}^2+q_{so}^2)} \nonumber\\
C^1_1&=&\frac{1}{D(q^2+q_{IR}^2+\frac{3}{2}q_{so}^2+\sqrt{4q^2q_{so}^2+q_{so}^4})}, \\
C^1_{-1}&=&\frac{1}{D(q^2+q_{IR}^2+\frac{3}{2}q_{so}^2-\sqrt{4q^2q_{so}^2+q_{so}^4})}. \nonumber
\ea
By integrating over momenta in Eq.\eqref{eq:UCF}, one finds 
\begin{widetext}
\ba
\delta\sigma^2\!&\! \sim \left \{
\begin{array}{ccc}
  \left(\frac{e^2}{h}\right)^2\!\left[4\!+\!\frac{1}{1\!+\!L^2q_B^2}\!+\!\frac{1}{1\!+\!L^2(q_B^2\!+\!q_{so}^2)}+F_1(L)+F_2(L)\right], & l_\phi\gg L \\
  \,\\
  \left(\frac{e^2}{h}\right)^2\!\frac{l_\phi^2}{L^2}\left[4\!+\!\frac{1}{1\!+\!l_\phi^2q_B^2}\!+\!\frac{1}{1\!+\!l_\phi^2(q_B^2\!+\!q_{so}^2)}+F_1(l_\phi)+F_2(l_\phi)\right],  & l_\phi \ll L 
\end{array}
\right.
\ea
where
\ba
F_1(l)&=&-\frac{1}{3(1+l^2(q_B^2+q_{so}^2))}+\frac{64}{3(16+l^2(16q_B^2+7q_{so}^2))}+\frac{64lq_{so}\arctan\left(\frac{3lq_{so}}{\sqrt{16+l^2(16q_B^2+7q_{so}^2)}}\right)}{(16+l^2(16q_B^2+7q_{so}^2))^{3/2}}, \nonumber\\
F_2(l)&=&\frac{1}{5(1+l^2(q_B^2+q_{so}^2))}+\frac{64}{5(16+l^2(16q_B^2+7q_{so}^2))}+\frac{64lq_{so}\arctan\left(\frac{5lq_{so}}{\sqrt{16+l^2(16q_B^2+7q_{so}^2)}}\right)}{(16+l^2(16q_B^2+7q_{so}^2))^{3/2}}. \nonumber
\ea
\end{widetext}

\begin{figure}[h]
\includegraphics[width=8cm]{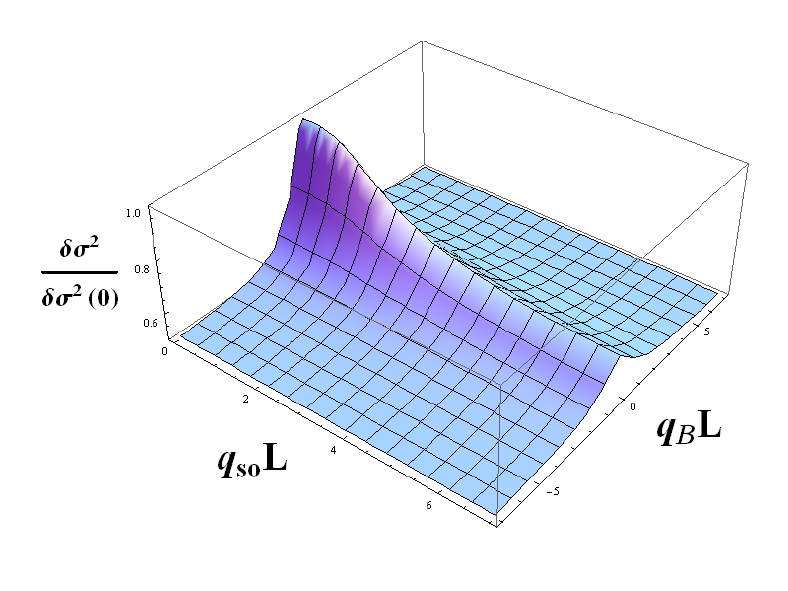}
\caption{Plot of UCF strength versus $q_{SO}$ and $q_B$ with linear Rashba SOI at $l_\phi \gg L$ limit.}
\label{fig:ucflinear}
\end{figure}
The dependence of UCF as a function of SOI and magnetic field is shown in Fig.~\ref{fig:ucflinear}. One can notice a suppression of $\delta\sigma^2$ by factor of $5/8$ for large SOI due to the suppression of the triplet contributions in the Cooper channel. The magnetic field suppresses the singlet Cooperon contribution, and $\delta\sigma^2/\delta\sigma^2(0)$ converges to 1/2 under strong magnetic field.

\subsubsection{UCF with cubic Rashba SOI}
Following similar steps as in the previous subsection, we compute Cooperon propagators with cubic Rashba SOI in a finite system:
\ba
\mathcal{C}^0_0&=&\frac{1}{D(q^2+q_{IR}^2)}, \,\,\,\,\,\,\,\,\,\,\,\,\,\,\, \mathcal{C}^1_0=\frac{1}{D(q^2+q_{IR}^2+2q_{so3}^2)}, \nonumber\\
\mathcal{C}^1_1&=&\frac{1}{D(q^2+q_{IR}^2+q_{so3}^2)}, \,\,\,\, \mathcal{C}^1_{-1}=\frac{1}{D(q^2+q_{IR}^2+q_{so3}^2)}.\nonumber
\ea
By integrating over momenta in Eq.\eqref{eq:UCF}, one obtains 
\begin{widetext}
\ba
\delta\sigma^2\!&\! \sim \left \{
\begin{array}{ccc}
 \left(\frac{e^2}{h}\right)^2\left[4+\frac{1}{1+L^2q_B^2}+\frac{2}{1+L^2(q_B^2+q_{so3}^2)}+\frac{1}{1+L^2(q_B^2+2q_{so3}^2)}\right], & l_\phi\gg L \\
  \,\\
  \left(\frac{e^2}{h}\right)^2\frac{l_\phi^2}{L^2}\left[4+\frac{1}{1+l_\phi^2q_B^2}+\frac{2}{1+l_\phi^2(q_B^2+q_{so3}^2)}+\frac{1}{1+l_\phi^2(q_B^2+2q_{so3}^2)}\right],  & l_\phi \ll L 
\end{array}
\right.
\ea
\end{widetext}

\begin{figure}[h]
\includegraphics[width=8cm]{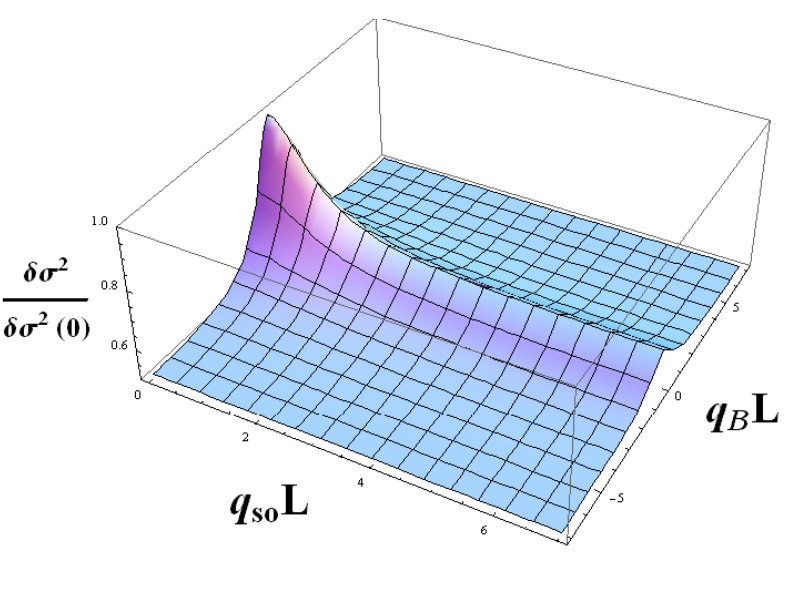}
\caption{Dependence of UCF on magnetic field and cubic Rashba SOI in $l_\phi \gg L$ limit.}
\label{fig:ucfcubic}
\end{figure}

The dependence of UCF on magnetic field and SO coupling is plotted in Fig. \ref{fig:ucfcubic}.
The suppression of the triplet channel contribution is steepr for cubic Rashba SOI than for linear Rashba coupling, as may be seen by comparing Fig.~\ref{fig:ucfcubic}
to Fig.~\ref{fig:ucflinear}.


\subsection{Quantum corrections to conductivity in quasi one-dimensional structures}

In this section, we consider quasi-one-dimensional system confined along the $y$-direction$(-\frac{W}{2}<y<\frac{W}{2})$. In this geometry, as previously discussed, the SOI in $d_{xy}$ and $d_{xz}$ bands at small carrier density is dominated by linear Rashba contribution. We, therefore, concentrate on this situation. Given the confinement along $y$-direction, we need to solve Eq.\eqref{eqn:ME} in real space and impose appropriate boundary conditions. The singlet component of the Cooperon is not affected by SOI, and, thus, the corresponding eigenvalue is $E^0$ taken at $q_y=0$. We now concentrate below on $J=1$ subspace. The matrix equation for the $J=1$ components of the Cooperon reads
\ba
&\hat{\mathcal{L}}_{J=1}(r)&\hat{C}_0(r,r')=\frac{1}{2\pi\hbar\nu\tau_0^2}\hat{\delta}(r,r')\\
&\hat{\mathcal{L}}_{J=1}(r)&=\frac{1}{\tau_\phi}+D\left[(-i\partial_x-q_{so}\hat{J}_y)^2+(-i\partial_y+q_{so}\hat{J}_x)^2  \right].\nonumber
\label{eqn:MEr}
\ea
Here $r=(x,y)$. The solution of above equation is given by,
\ba
&\hat{C}&(r,r')=\frac{1}{2\pi\hbar\nu\tau_0^2}\sum_{m=-1}^{1}\frac{\ket{\psi_m(r)}\bra{\psi_m(r')}}{E_m}\\
&\hat{\mathcal{L}}_{J=1}&(r)\ket{\psi_m(r)}=E_m\ket{\psi_m(r)}
\ea
with boundary conditions,
\be
(-i\partial_y+q_{so}\hat{J}_x)\ket{\psi_m(r)}|_{y=\pm \frac{W}{2}}=0
\ee
implying zero current in the direction normal to the boundary for each spin eigenstates. We note that $q_y=0$ does not satisfy the boundary conditions. To find the Cooperon propagator in this case, we first perform a gauge transformation\cite{Aleiner2001, Kettemann07} and simplify boundary conditions. Let's perform the canonical transformation defined by $\hat{U}(y)=e^{iq_{so}\hat{J}_x y}$ and introduce  $\ket{\tilde{\psi}_m(r)}=U(y)\ket{\psi_m(r)}$ and $\tilde{\mathcal{L}}(r)=\hat{U}(y)\hat{\mathcal{L}}_{J=1}(r)\hat{U}^\dagger(y)$ where  $\tilde{\mathcal{L}}$
\ba
\tilde{\mathcal{L}}(r)&=&\frac{1}{\tau_\phi}+D[(-i\partial_x)^2+(-i\partial_y)^2]\nonumber\\
&&-2Dq_{so}[e^{iq_{so}\hat{J}_x y} \hat{J}_y e^{-iq_{so}\hat{J}_x y}](-i\partial_x)\nonumber\\
&&+Dq_{so}^2[e^{iq_{so}\hat{J}_x y} \hat{J}_y^2 e^{-iq_{so}\hat{J}_x y}].
\ea
In terms of the new eigenstates, the boundary condition reads
\ba
(-i&\partial_y)&\ket{\tilde{\psi}_m(r)}|_{y=\pm \frac{W}{2}}=0
\ea
and, thus, the zero mode $q_y=0$ now satisfies the above boundary condition. If the width satisfies $W\ll 1/q_\phi$ as $1/q_\phi$ being dephasing length, one can neglect higher harmonics, $n_y \geq 1$, because they are suppressed by a factor of $W q_\phi$. In this regime, the dominant contribution comes from $q_y \propto n_y =0$ mode.
Furthermore, in the limit $W\ll 1/q_{so}$, $\tilde{\mathcal{L}}(r)$ is a slowly varying function of $y$, and can be approximated by its average over $\hat y$-direction. Then, we find that
\begin{align}
&\frac{\tilde{L}(q_x,0)}{D}=\frac{1}{DW}\int_{-W/2}^{W/2} dy\tilde{L}(q_x,y)=q_\phi^2+q_x^2+\hat{G}_1+\hat{G}_2,\nonumber\\
&\!\hat{G}_1\!=\!-2q_{so}q_x\frac{2\sin(\frac{q_{so}W}{2})}{q_{so}W}\hat{J}_y,\nonumber\\
&\!\hat{G}_2\!=\!\frac{q_{so}^2}{4}\!\left(\!\!
\begin{array}{ccc}
 3\!-\!\frac{\sin(q_{so} W)}{q_{so}W} & 0 & -1\!+\!\frac{\sin(q_{so} W)}{q_{so}W}\\
 0& 2 \!+\! 2\frac{\sin(q_{so} W)}{q_{so}W} & 0 \\
 -1\!+\!\frac{\sin(q_{so} W)}{q_{so}W} & 0 &  3\!-\!\frac{\sin(q_{so} W)}{q_{so}W}
\end{array}
\!\!\right).\nonumber
\end{align}
The eigenvalues of $\tilde{L}(q_x,0)/D$ are given by
\ba
E^1_0/D&=&q_\phi^2+q_x^2+\frac{q_{so}^2}{2}t_{so},\nonumber\\
E^1_\pm/D&=&q_\phi^2+q_x^2+\frac{q_{so}^2}{4}\left(4-t_{so}\pm\sqrt{t_{so}^2+\frac{64q_x^2}{q_{so}^2}(1-c_{so})^2} \right),\nonumber\\
t_{so}&=&1-\frac{\sin(q_{so}W)}{q_{so}W}\sim \frac{(q_{so}W)^2}{6}, \nonumber\\
c_{so}&=&1-\frac{2\sin(\frac{q_{so}W}{2})}{q_{so}W}\sim \frac{(q_{so}W)^2}{24}.\nonumber
\ea
With the Cooperon propagator in hand, we can now
compute quantum corrections to the conductivity
\be
\Delta\sigma_{1D}=-\frac{2e^2D}{h}\int_{q_{\rm min}}^{q_{\rm max}}\frac{dq_x}{2\pi}\left(-\frac{1}{E^0}+\sum_{m=-1}^1\frac{1}{E^1_m}\right)
\ee
At non-zero magnetic field, this expression is modifed by introducing an additional cutoff $q_B^2$:
\ba
\Delta\sigma_{1D}(B)&=&-\frac{e^2}{h}\bigg[-\frac{1}{\sqrt{q_\phi^2+q_B^2}}+ \frac{1}{\sqrt{q_\phi^2+q_B^2+2rq_{so}^2}}\nonumber\\
 &&+\frac{2}{\sqrt{q_\phi^2+q_B^2+rq_{so}^2}}\bigg],
\ea
where $r$ is a width-dependent coefficient that characterizes the effective strength of the spin-orbit coupling, $r=(q_{so}W)^2/12$. The magnetic field cutoff is also modified and becomes width-dependent. For weak fields $B \ll h/e W^2 $, the cutoff $q_B^2\sim e^2 B^2 W^2/h^2$ whereas for large fields $B \gg h/e W^2 $, it remains the same as in 2D, {\it i.e.} $q_B^2 \sim eB/h$.
%
\\

Throughout this section, we have assumed that the system is in the diffusive regime
(i.e. all lengths are longer than the elastic mean free path) and have derived the Cooperon propagator in a quasi-1D system in this limit, i.e. assuming that $q^{-1}_{so}\gg  W \gg l_e$, where $l_e$ is the mean-free path. The results of our calculation can be extended to a quasi-1D nanowire whose width is comparable with the mean free path $l_e$; see, for example, Ref.~\onlinecite{Beenakker88}. In this regime, the magnetic field cutoff should be modified due to the flux cancellation effect. In the weak ($B\ll\hbar/eWl_e$) and strong ($B\gg \hbar/eWl_e$) magnetic field limit, the cutoff $q_B$ becomes $q_B^2=2e^2B^2W^3/(C_1\hbar^2 l_e)$ and $q_B^2=2eBW^2/(C_2\hbar l_e^2)$, respectively. Here, the coefficients $C_1=9.5$ and $C_2=4.8$ are obtained for specular boundary condition.


\subsection{Universal Conductance Fluctuations in a Quasi-One Dimensional System}
\begin{figure}[h]
\includegraphics[width=8cm]{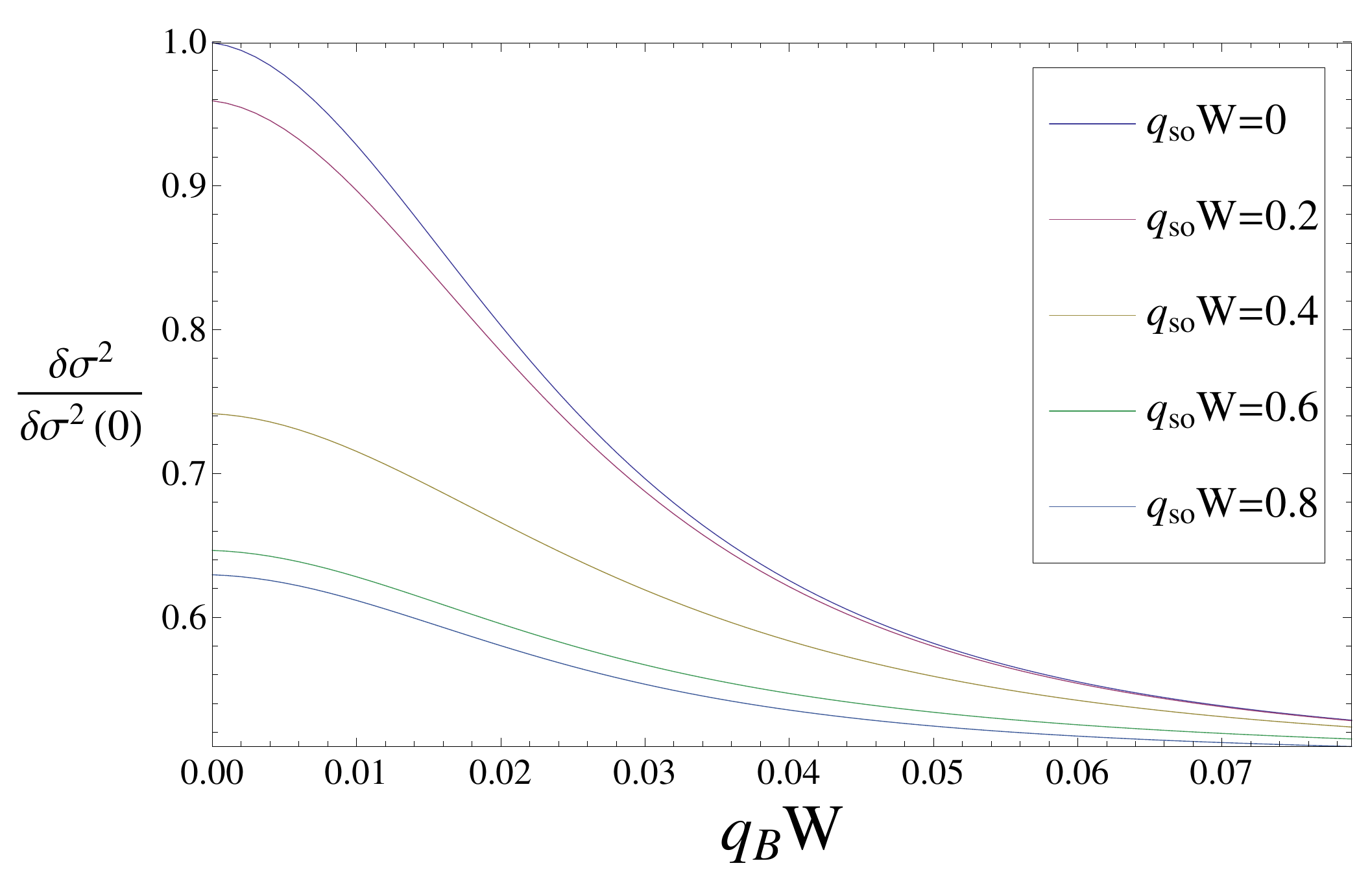}
\caption{Plot of $\delta\sigma^2/\delta\sigma^2(0)$ versus magnetic field for various SOI strength at $L=30W$.}
\label{fig:ucf1d}
\end{figure}

Using the expressions for Cooperon propagators derived in the previous section, we now evaluate UCF in a quasi-one dimensional system with width $W\ll 1/q_\phi$, $1/q_{so}$ and $L\gg l_\phi$. For a mesoscopic system with length $L\ll l_\phi$, we need to change $q_\phi \rightarrow 1/L$ in the expressions for the Cooperon propagator. Then, we find that the quasi one-dimensional conductivity variance is given by
\be
\delta\sigma_{1D}^2\sim \left(\frac{e^2}{h}\right)^2\cdot\frac{D^2}{L}\int^{q_{max}}_{q_{min}}\frac{dq}{2\pi}\left({\rm Tr}[\hat{D}(q)]^2 + {\rm Tr} [\hat{C}(q)]^2\right)
\ee
We evaluate the momentum integral numerically, assuming a fixed ratio of $W$ and $L$.
The dependence of the variance of the conductivity on SO coupling and magnetic field for $L=30W$ is shown in Fig. \ref{fig:ucf1d}.
We find that UCF in a quasi-1D system depends on the magnetic field in a manner similar to its dependence in 2D, i.e. $\delta\sigma_{1D}^2/\delta\sigma_{1D}^2(0)$ converges to the same value 1/2 under strong magnetic field.
\section{Discussion of Experimental Results}\label{sec:exp}

\subsection{Weak Anti-Localization Measurements at STO Surfaces}

Nakamura {\it et al}. \cite{Nakamura'12} recently reported evidence of a cubic Rashba SOI at a low carrier density ($k_Fa<0.3$) STO surface. They concluded that the bottom $d_{xy}$-like band has cubic Rashba SOI. However, as may be seen in Fig. \ref{fig:a1STO}, we found that the SOI of bottom $d_{xy}$-like band is dominated by linear Rashba SOI in the region of small carrier density probed in the experiment.
\begin{figure}[h]
\includegraphics[width=8cm]{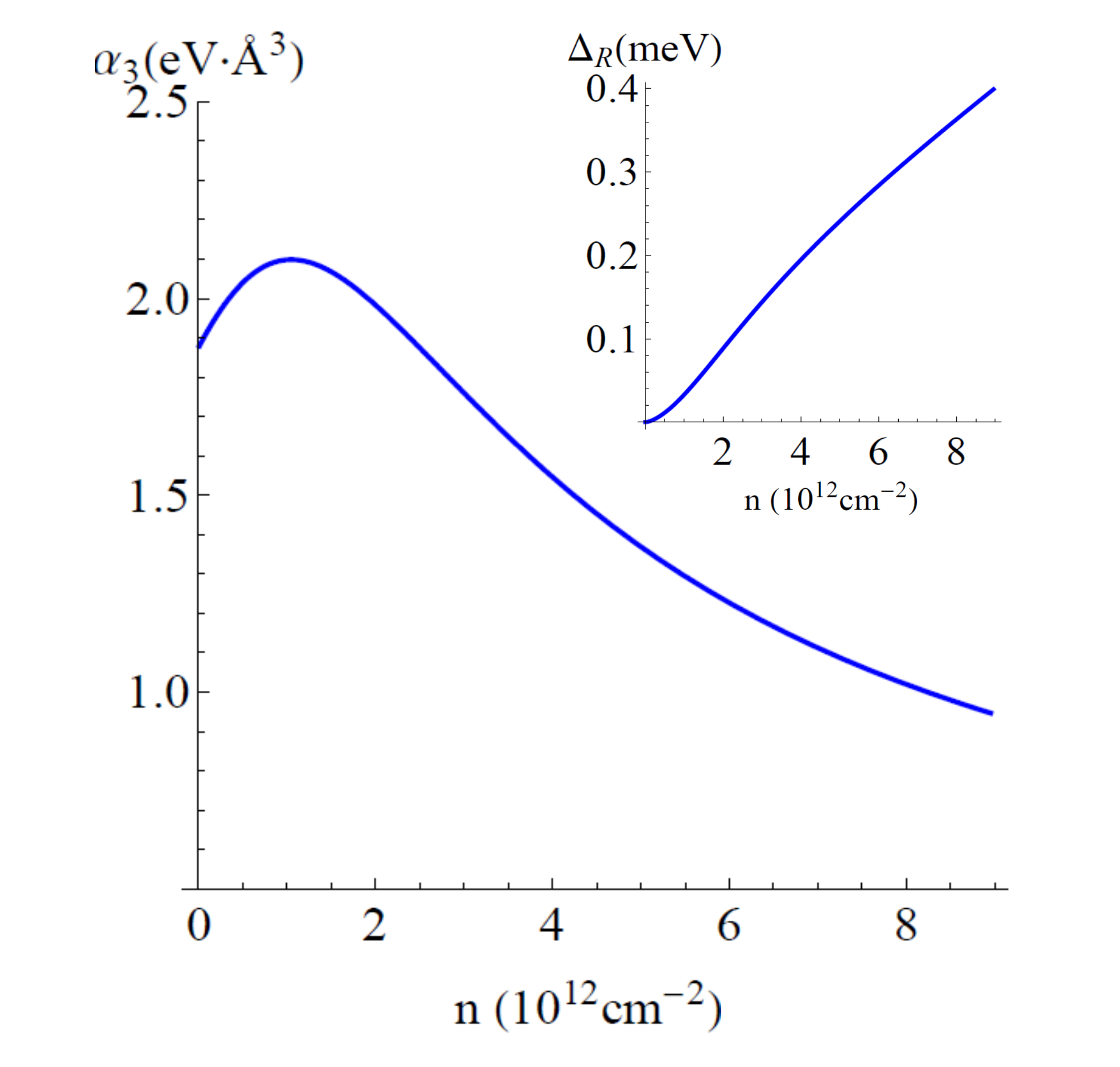}
\caption{Dependence of spin-orbit coupling $\alpha_3\,$(eV$\cdot\AA^3$) and spin-orbit energy $\Delta_R\,$(meV) on electron density $n\,$($\times10^{12}$cm$^{-2}$) in the middle band of STO surfaces. For energy spectrum see Fig.~\ref{fig:hrband}. Here we used $\Delta_z=35\,$meV, $\Delta_{ASO}=6\,$meV. \label{fig:a3STOm}}
\end{figure}

Since this is in contradiction with the data of Nakamura {\it et al}. \cite{Nakamura'12}, we assume, instead, that electrons in the bottom band are localized (or very poorly conducting) and that the observed transport is due to the middle band. In that case, WAL should be due to a cubic Rashba term since the middle band has only cubic Rashba SOI. The calculated values for the cubic Rashba coupling coefficient as a function of carrier density are plotted in Fig. \ref{fig:a3STOm}.


\subsection{Weak Anti-Localization Measurements at LAO/STO Interface}

\begin{figure}[h]
\includegraphics[width=8cm]{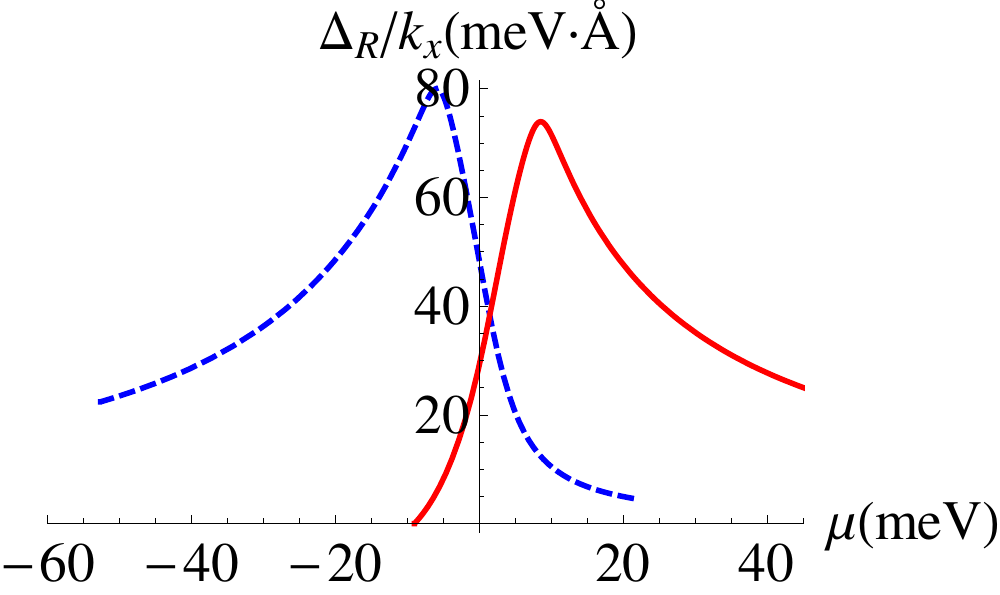}
\caption{Calculated $\alpha_1$($\Delta_R/k_x$) vs $\mu$(meV). Dashed (solid) plots correspond to bottom (middle) band. \label{fig:a1LAOSTO}}
\end{figure}

We now discuss SOI using our model for LAO/STO interfaces and apply it to explain rapidly increasing Rashba SOI observed in recent experiments~\cite{Caviglia10,BenShalom,Triscone12}. We consider a three-band model, as we did for the surface of STO, and take the values $\Delta_E=50\,$meV from x-ray absorption spectroscopy \cite{Salluzzo09} and $\Delta_{ASO}=9\,$meV and $\Delta_z=20\,$meV from the DFT calculations of Ref. \onlinecite{Zhong12}. The dependence of the strength of SOI, following from the three-band model, on chemical potential is shown in Fig. \ref{fig:a1LAOSTO}. Dashed and solid curves correspond, respectively, to bottom (linear Rashba SOI) and middle (cubic Rashba SOI) bands. Comparison our results with the experiment provides two possible explanations for the rapid increase of SOI at a specific gate voltage.
The first hypothesis is that the transport at that gate voltage is due to the bottom $d_{xy}$-like band. This band has linear Rashba SOI at $k\approx 0$ and a much larger SOI near the avoided crossing with the $d_{xz,yz}$ bands. Approaching the crossing causes reduction of the gap, which, in turn, leads to an enhancement of SOI. For this hypothesis to agree with the observed data \cite{Caviglia10}, superconductivity would have to be suppressed once electrons start populating the middle band since the superconducting transition temperature starts to decrease right after the sharp increase of SOI. It is not clear why this should be the case. Furthermore, the observed carrier density of the bottom band, as deduced by Hall measurements, $\sim 10^{13}\,$cm$^{-2}$ seems to be much smaller than the polar catastrophe theory suggests $\sim10^{14}\,$cm$^{-2}$. Moreover, it is not easy to see why there should be a sharp onset density at which superconductivity starts to appear.

The other hypothesis is that these observations are dominated by transport from the middle heavy electron band. We now apply  our four-band model, and assume that most of the $\sim 10^{14}\,$cm$^{-2}$ electrons predicted by the polar catastrophe argument are localized at the interface.
A much smaller number $\sim10^{13}\,$cm$^{-2}$ of electrons populates the middle band and dominates transport. Electrons in the second $d_{xy}$ sub-band may contribute to the Hall effect, but the WAL phenomena seen in experiments would be due to cubic Rashba for small mobile carrier density, as at the STO surfaces. This picture also suggests that superconductivity arises as a result of the appearance of electrons in the middle band.

Yet another possibility to consider is transport in quasi-one-dimensional channels, as in wires `drawn'
with an AFM tip \cite{Cen08}, or appearing spontaneously and are related to the formation of tetragonal domains formed in the STO below 105$\,$K~\cite{Kalisky2013}. As we discussed above, Rashba SOI is always linear in momentum in quasi-1D geometry, regardless of the band index.  There is, however, an important feature of transport in the quasi-$1D$ case that does distinguish between
carriers in the different bands. At the values of the chemical potential at
which the $d_{xz}$ and $d_{yz}$ bands become degenerate, the spin-orbit interaction in these bands vanishes, as shown in Fig. \ref{fig:1d}.
Therefore, it would be interesting to revisit WAL data and fit the magnetoconductance having a particular scenario in mind.


\section{Conclusions}

The origin of the physics underlying the ordering phenomena of LAO/STO interfaces,
namely magnetism and superconductivity, is still unclear and controversial.
However, spin-orbit interaction may provide a window into understanding
these properties. As we have seen, the electrons in the different
bands of the LAO/STO interface have Rashba
(i.e. interface-induced) SOI with different momentum dependences. Moreover,
this dependence is a strong function of the effective dimensionality
of the carriers. The momentum-dependence and effective dimension
are, in turn, reflected in transport measurements through the dependence
of WAL effects on an applied magnetic field. In this paper,
we have given a simple explanation for both the nature of
SOI in LAO/STO and the surface of STO and also for its
WAL signature.

We find that the sharp increase in the strength of the SOI
with gate voltage is consistent with conduction that is dominated
by either the $d_{xy}$ or $d_{xz,yz}$ bands. However, the
spin-orbit energy scale is predicted to decrease at still higher
gate voltages. This decrease would steeper in the case of the
$d_{xy}$ band. Therefore, if it were possible to increase the gate
voltage until the spin-orbit coupling peaks and begins to decrease
(as the superconducting transition is observed to do), it would be possible
to distinguish between these two scenarios. In the quasi-$1D$ case, the
difference between the $d_{xy}$ and $d_{xz,yz}$ bands may be even more dramatic
since the SOI vanishes in the latter case
at one value of the chemical potential. If nominally 2D transport is
actually quasi-$1D$, as suggested by recent measurements\cite{Kalisky2013},
then there may be a third possible functional
form against which WAL data on LAO/STO interfaces
could be measured. However, there are, at present, too many unknowns
(such as the wire width and spacing) to make a meaningful comparison
between theory and experiment. Finally, we note that we have made concrete predictions for the dependence of universal conductance fluctuations on the mangetic field, spin-orbit interaction, and device size, which could be compared to experiments if the dependence on these parameters coud be measured in experiments similar to those reported recently in Ref.~\onlinecite{Stornaiuolo13}.

\acknowledgments{We would like to thank Andrea Caviglia, Guanglei Cheng, Harold Hwang,
Jeremy Levy, Kam Moler, Susanne Stemmer, Minseok Choi and Joshua Veazey for discussions.
C.N. is supported by the DARPA QuEST program and the AFOSR under grant FA9550-10-1-
0524. Y.K. is supported by the Samsung scholarship and the Microsoft Research Station Q. C.N. and R.M.L. thank the Aspen Center for Physics for hospitality and support under NSF grant \#1066293.}

\bibliography{LAO-STO}

\end{document}